\documentclass{PoS}
\usepackage{colordvi}
\usepackage{graphicx}
\graphicspath{{./Figures/}}

\leftline{\parbox{3cm}{\large\rm HU-EP-09/62}
\vspace {1.0cm}}

\title{
The Landau gauge gluon propagator in 4D SU(2) lattice gauge theory 
revisited: Gribov copies and scaling properties
}
\ShortTitle{Gluon propagator in SU2 LGT revisited}
\author{\speaker{I.~L.~Bogolubsky}$^{\hspace{1mm}a}$,
  E.-M.~Ilgenfritz$^b$, M.~M\"uller-Preussker$^{b}$, and A.~Sternbeck$^{c}$\\~\\
  $^{a}$ Joint Institute for Nuclear Research, 141980 Dubna, Russia\\
  $^{b}$ Humboldt Universit\"at zu Berlin, Institut f\"ur Physik, 12489 Berlin,
  Germany\\
  $^{c}$ CSSM, School of Chemistry \& Physics, University of Adelaide,
  SA 5005, Australia\\
  E-mail: \email{bogolubs@jinr.ru}, \email{ilgenfri@physik.hu-berlin.de},
  \email{mmp@physik.hu-berlin.de},
  \email{andre.sternbeck@physik.uni-regensburg.de}
}
\date{December 5, 2009}

\abstract{Lattice results for the gluon propagator in $SU(2)$ 
 pure gauge theory obtained on large lattices are presented.
 Simulated annealing is used throughout to fix the Landau gauge.
 We concentrate on checks for Gribov copy effects and for scaling
 properties. Our findings are similar to the ones in the $SU(3)$ 
 case, supporting the decoupling-type infrared behaviour of 
 the gluon propagator.}

\FullConference{The XXVII International Symposium 
 on Lattice Field Theory - LAT2009 \\ July 26-31 2009\\
 Peking University, Beijing, China}

\begin{document}

\vspace{-0.3cm}
\section{Introduction}
%---------------------
\vspace{-0.1cm}

Over the last years interesting results in Landau gauge 
gluodynamics have been found for gluon and ghost propagators 
(and, consequently, for the running coupling)
both within the semianalytical Dyson-Schwinger(DS) as well as 
Functional Renormalization Group (FRG) 
approaches ~\cite{ORSAY08, FMP08} and with the help of 
lattice computations~\cite{CM07, CM08, SSLW07, WE07, WE09}.
They have excited controversal debates about the behaviour in the 
deep infrared (IR) region. Within the DS and FRG approach it
was demonstrated \cite{FMP08} that the behaviour strongly depends 
on the choice of the infrared limit of the ghost dressing function
taken as a boundary condition for solving the (truncated) system
of equations. The so-called {\it scaling solution} exhibits
an IR singular well-defined power-like behaviour of the ghost 
dressing function and correspondingly a vanishing gluon 
propagator in agreement with the quite attractive confinement 
scenarios invented some time ago by Gribov and Zwanziger on one hand
and by Kugo and Ojima on the other. Moreover, it was in accordance with 
BRST invariance properties. 

Lattice results -- as long as they are based on the assumption that one 
has to choose Gribov gauge copies as close as possible to the 
{\it fundamental modular region} -- support with convincing numerical 
evidence the so-called {\it decoupling solution} with finite IR 
limits of both the gluon propagator and the ghost dressing function. 
Revisiting the case of $SU(2)$ lattice gauge theory we give here
further evidence for this observation by a consequent use of an
efficient gauge fixing method, the {\it simulated annealing} algorithm.
Since previous lattice investigations of the IR limit both in $SU(2)$ 
as well as in $SU(3)$ were carried out at quite strong bare coupling
values in order to reach largest possible physical volumes the
continuum limit was not really under control. Therefore, in the given
contribution to LATTICE '09 we have a look into the scaling properties
of the gluon propagator, which seem to be a bit more involved in the 
$SU(2)$ than in the $SU(3)$ case. We neglect so-called $SU(2)$-flips 
which enlarge the class of Landau gauge orbits and allow to extremize 
the gauge functional even further.

\section{Landau gauge fixing, Gribov ambiguity and gluon propagator}
%-------------------------------------------------------------------
\vspace{-0.1cm}

In order to fix the gauge on the lattice we apply gauge transformations 
$~g(x) \in G=SU(N_c)$, $(N_c=2,3)~$ to sets $~\{U_{x,\mu}\}~$ 
of link variables by mapping 
$~U_{x,\mu} \to {}^gU_{x,\mu}=g(x) U_{x,\mu} g^{\dagger}(x+\hat{\mu})\,.$
The set of all admissible $~\{{}^gU_{x,\mu}\}~$ for a given field 
$~\{U_{x,\mu}\}~$ is called a gauge orbit. The Landau gauge 
$~\partial_\mu A_\mu =0~$ for 
$~A_{x+\hat{\mu}/2,\mu} = (1 / 2\, i\, a\, g_0)
  (U_{x,\mu}-U^{\dagger}_{x,\mu} )_{\rm traceless}~$
is fixed by searching for the local maxima $~g_{lmx}(x)~$ of the gauge 
functional 
\begin{equation}
F_U[g] = \frac{1}{N_c} \sum_{x,\mu} \mathfrak{Re}~\mathrm{Tr}~{}^gU_{x,\mu} \;.
 \label{fu}
\end{equation}
In general for non-Abelian groups $G$ more than one local maximum
$~g_{lmx}(x)~$ can be found, the so-called Gribov copies. Since the values of
the gauge functional $~F_U[g]~$ and other gauge-variant quantities $~O({}^gU)~$
computed for various Gribov copies typically are correlated, further
clarification of the gauge fixing condition is required. In case of
Landau (or Coulomb) gauges for $SU(N_c)$ gauge theories it was
proposed~\cite{Zwan94} to choose the {\it global} maximum 
$~g_{gmx}(x)~$ among all local ones, thus introducing the 
{\it fundamental modular region} (FMR) inside the {\it Gribov region}, 
the latter defined by the positivity region of the Faddeev-Popov 
operator~\cite{Zwan94}. In the present study we still keep
the FMR condition. In practice it is hard - if not even impossible -
to reach the FMR. Therefore, one is interested to improve the gauge
fixing method 
and/or to apply the method of choice many times starting from random 
gauges in order to find the ``best copies'' closest to the FMR.

In what follows we reconsider the question of how the gluon
propagator depends on the Gribov ambiguity when the (Landau)
gauge is fixed on a lattice. This question has been addressed 
for relatively small lattice volumes already in preceding
publications~(e.g. in \cite{Cu97, BIMMP, SO04}). Note that in 
previous papers, instead of a comparison between two different 
gauge-fixing techniques, a comparison is made of the ``best copy'' 
(bc) with respect to the maximal $F_U[g]$ value achieved and ``first copy''  
(fc), i.e. corresponding to a randomly chosen copy~\cite{BIMMP, SIMPS}. 
In ~\cite{SO04} ``worst'' copies were used for comparison, as well. 
There a visible Gribov copy effect was reported for the gluon propagator
in the infrared. However, the effect appeared to be much more pronounced, 
when (i) the gauge orbit was extended by admitting 
nonperiodic (periodic up to $Z(N_c)$) gauge transformations ({\it flips}) 
and when (ii) the standard overrelaxation method (OR) was replaced by a 
{\it simulated annealing} algorithm (SA) - always followed by finalizing OR 
steps~\cite{BBMMP, BBBIMPM, BMMP08}. Moreover, the more efficient
gauge fixing approach (SA + flips) led to a suppression of finite-volume 
effects, and indications were found that the influence of flip 
transformations on the gluon propagator gradually weakens with 
increasing linear lattice size $L$. But all these results were 
obtained at rather small lattice volumes.

In the present study for $SU(2)$ we neglect the flip gauge transformations
and compare the SA method with the OR algorithm for strictly periodic 
gauge transformations. Our implementation of SA gauge fixing in the 
$SU(2)$ case differs from that for $SU(3)$ gluodynamics~\cite{WE07,WE09} 
only in some technical details. We note that the ``temperature'' $T_{max}$,
from which SA cooling starts in the $SU(2)$ case, is chosen to be 
the ``critical'' value $T_{cr}$ of some phase transformation
~\cite{SCH, WE09}, which takes place in the gauge-fixing field
$g(x)$ interacting with the Monte Carlo generated equilibrium field 
$U_{x,\mu}$ according to the gauge functional (\ref{fu}).
For the $SU(2)$ case it looks like a higher order phase
transition \cite{SCH} or even like a ``crossover''. Anyhow, 
we have chosen $T_{max}=T_{cr}=1.1$ in most cases.
Simulated annealing, also known as ``stochastic optimization
method''~\cite{SA12}, in principle allows getting arbitrarily close 
to the global maximum $F_U[g_{gmx}]$, if the number of 
SA ``cooling steps'' is large enough. This is the
underlying idea of our ``single copy'' method successfully used
in our $SU(3)$ papers~\cite{WE07, WE09}. In fact, in our $SU(2)$ 
computations we have used long SA chains with $O(10^4)$ 
steps from $T_{\rm max}$ down to $T_{\rm min}=0.01$~\cite{WE07}.

\begin{figure}
\begin{center}
\includegraphics[height=7.6cm,width=7.8cm,angle=270]{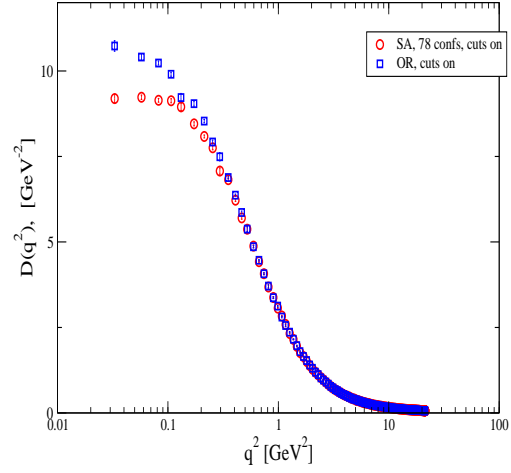} 
\hspace{-0.3cm}
\end{center}
\caption{Comparison of the unrenormalized gluon 
propagator obtained with SA and OR gauge fixing methods. The OR data
are taken from Ref. \cite{SSLW07}.}
\label{fig:D80_SAvsOR}
\end{figure}

The comparison for the unrenormalized gluon propagator obtained with 
SA versus OR techniques on $80^4$ lattices at $\beta=2.30$ is shown
in Fig.~\ref{fig:D80_SAvsOR}, where we have plotted only results 
for momenta $q^2$ surviving the so-called cylinder cut~\cite{LSWP,SSLW07}. 
One can clearly see a noticeable difference between SA and OR gluon 
propagator values in the deep infrared region $q^2 < 0.2 GeV^2$, 
where the Gribov effect leads to a qualitative change of the behaviour
of the gluon propagator. The question remains, whether 
the Gribov copy effect for the gluon propagator weakens with 
a further increase of the lattice volume. 

\section{Check of scaling and multiplicative renormalizability}
%----------------------------------------------------------------

The progress to reach the infrared regime had the price to 
consider lattice gauge fields on rather coarse lattices. To our
knowledge the continuum limit expressed in a proper scaling 
and a multiplicative renormalization behaviour was not yet  
considered in detail on large volumes. We have made a step 
into this direction again neglecting the influence of $Z(2)$ flip
transformations.

\begin{figure}
\centering
\includegraphics[height=8.0cm,angle=270]{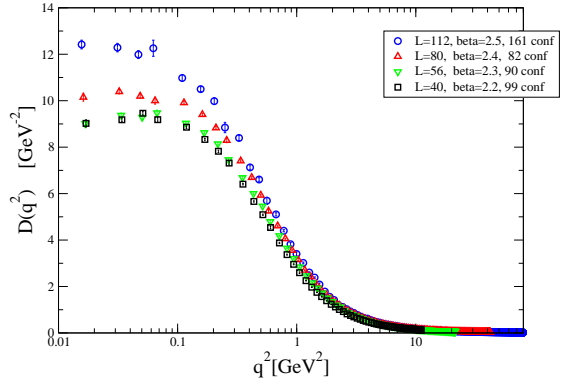}%
\caption{The unrenormalized gluon propagator $D(q^2)$ for various 
($L,\beta$) pairs, i.e. at fixed physical volume.}
\label{fig:D_varLbet_McRn}
\end{figure}

We have computed the gluon propagator on a sequence of lattices 
with increasing linear lattice size $L$ and $\beta$, choosing 
($L$, $\beta$) such that the physical volume was kept more or 
less constant. We have produced equilibrium ensembles of MC 
configurations and fixed the Landau gauge with the 
(single-copy) SA method for: 
$(L, \beta) = (40, 2.2), (56, 2.3), (80, 2.4), (112,2.5)$, 
i.e. for a physical box size of approximately $10 \mathrm{~fm}$.
The bare gluon propagator $D(q^2)$ for these pairs
of parameters is shown in Fig.~\ref{fig:D_varLbet_McRn}. 

There are quite strong differences which have not been
observed in the $SU(3)$ case before. But this does not come 
unexpected. In the continuum or scaling limit the bare propagator 
is expected to be multiplicative renormalizable as in perturbation theory. 
This means that between the results obtained at different lattice 
cutoffs a finite multiplicative renormalization up to lattice
artifacts should be possible. In a finite volume 
-- unavoidable for any lattice results -- the multiplicative 
renormalization could be violated by finite-size effects.  

In accordance with the standard momentum-subtraction (MOM) 
renormalization scheme we have multiplicatively rescaled the 
bare gluon dressing functions $Z(q^2, \beta) \equiv q^2 D(q^2)$ 
(for all $\beta$ values considered) equating the renormalized 
values at some $q=\mu$ to the tree-level value $Z_{ren}(\mu^2) = 1$.
The renormalization point was chosen at $\mu^2 = 5.8 \mathrm{~GeV}^2$,
sufficiently far away from the cutoff momentum 
$q_{max}^2= 79.4 \mathrm{~GeV}^2$ for $\beta=2.50$. For illustration
the finite renormalization factors $Z(\mu^2, \beta) / Z(\mu^2, \beta=2.5)$
are shown in the Table~\ref{tab:table1}. 
They were obtained by interpolating between 
the 7 data points closest to the chosen scale $\mu$ for $\beta=2.5$. 

\begin{table}[ht]
\caption{Finite renormalization factors for $\mu^2 = 5.8 \mathrm{GeV}^2$.}
\centering
\begin{tabular}{|c|c|}
   \hline
$\beta$&$Z(\mu^2,\beta)/Z(\mu^2, \beta=2.5)$ \\
   \hline
2.2  &  0.815  \\
2.3  &  0.8925 \\
2.4  &  0.9489 \\
\hline
\end{tabular}
\label{tab:table1}
\end{table}

\begin{figure}
\centering
\includegraphics[height=8.0cm,angle=270]{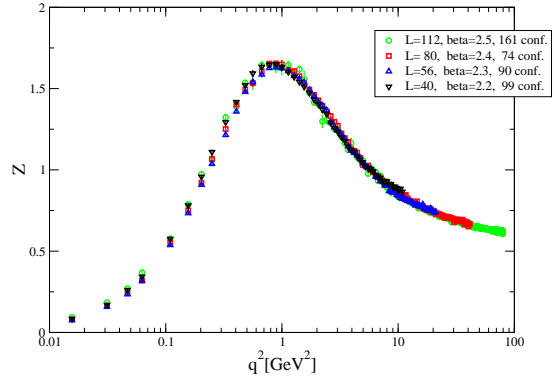}%
\caption{The renormalized gluon dressing function in the MOM scheme.}
\label{fig:Z_McRn}
\end{figure}

The MOM-renormalized dressing function is plotted in Fig.~\ref{fig:Z_McRn}.
One clearly sees that the three curves for the renormalized dressing function
$Z_{ren}(q^2)$ obtained at lattice sizes $L=56, 80, 112$ lie nicely on top 
of each other, thus confirming the expected multiplicative 
renormalizability. For $\beta=2.2$ and $L=40$ there are some scaling 
violations that can be understood as lattice artifacts.
Some slight variations of the curve obtained with the largest lattice
$L=112$ probably can be attributed to problems with the still unsufficient
Monte Carlo statistics and/or autocorrelations. 

\section{ Conclusions}
%---------------------

The comparison of OR- and SA-based results for the gluon propagator 
for $L=80$ and $\beta=2.3$ clearly shows a noticeable Gribov copy 
effect in the range of momenta  $q^2 < 0.2 GeV^2$. 
At the moment the (dis)appearance of this effect for even larger
volumes is an interesting open problem. 
An open question is also, whether our results 
obtained on large volumes will be modified, when $Z(2)$ flips 
are taken into account. This is a matter of research in a
forthcoming papers of one of the coauthors~\cite{BMMP09}. 

Using SA-based Landau gauge fixing we have got numerical confirmation 
for a nonperturbative multiplicative renormalizability for the 
gluon dressing function. Finite-size effects already seem to be 
negligible in the given range of momenta.

\end{document}